\documentclass{elsart41}
\usepackage{graphics}
\usepackage{graphicx}
\usepackage{amssymb}
\begin{document}

\begin{frontmatter}


\title{Local moment approach to multi-orbital single impurity Anderson 
  model; application to dynamical mean-field theory}
\author{Anna Kauch\corauthref{cor1}},
\author{Krzysztof Byczuk}

\address{%
Institute of Theoretical Physics,
Warsaw University, 
ul. Ho\.za 69,
 PL-00-681 Warszawa,
 Poland
}

\corauth[cor1]{%
Corresponding author.\\  Tel:+48 (22) 5532315
fax: +48 (22) 6219 475 \\
\textit{Email address:} anna.kauch@fuw.edu.pl (A. Kauch)}

\begin{abstract}
Using a local moment approach of Logan {\it et al.} we
developed a solver for a multi-orbital single impurity Anderson model.
The existence of the local moments is taken from the outset and their values are
determined through variational principle by minimizing the corresponding
ground state energy.
The method is used to solve the dynamical mean-field
  equations for the
  multi-orbital Hubbard  model. In particular, the Mott-Hubbard
  metal--insulator transition is addressed within this approach.

\end{abstract}


\end{frontmatter}


The problem of a magnetic impurity in a metallic host, i.e. the Kondo problem,
 plays a major role in many branches of modern condensed matter physics.
This problem of coupling degenerate spin degrees of freedom to  gapless fermionic
excitations is essential in the dynamical mean--field theory (DMFT) of strongly
correlated electron systems \cite{vollhardt}.
The DMFT is a self--consistent method, where in each iteration loop the Kondo
problem is solved numerically. 
The application of the DMFT to investigate real systems, in particular with
many orbitals, is difficult because of the necessity to solve 
the Kondo problem for arbitrary parameters. 
Numerically exact methods, like numerical renormalization group \cite{bulla} or determinant
quantum Monte Carlo \cite{georges} are very
time (CPU) consuming, in particular when the number of orbitals is large. 
Reliable analytical methods are therefore needed. 
One of such methods, which recovers Luttinger property and Kondo exponential
scale as well as Hubbard satellites has been recently invented by Logan {\it et
  al}. \cite{Logan} and named a {\it local moment approach} (LMA). 
In this contribution we generalize this method to investigate multi-orbital
systems at zero temperature. 
Within the DMFT framework supplemented with the
LMA we analyze the problem of Mott--Hubbard metal--insulator transition (MIT)
 in orbitally degenerate and non--degenerate systems.

We start with the generalization of the LMA to solve the single
impurity Anderson model (SIAM) with many orbital levels:
\begin{eqnarray}
H_{\rm{SIAM}}=\sum_{\alpha,\sigma}\left(\epsilon_{\alpha}-U_{\alpha}
n_{\alpha,\bar{\sigma}}\right) n_{\alpha,\sigma}+ 
\nonumber \\
\sum_{\sigma,\sigma'}\sum_{\alpha\neq\beta}\left( U'_{\alpha\beta}
-J\delta_{\sigma\sigma'}\right)
n_{\alpha\sigma}n_{\beta\sigma'}+ \\
\sum_{{\bf k},\sigma,\alpha} V_{{\bf k}\alpha}\left(
d_{\alpha\sigma}^{\dagger} c_{{\bf k}\sigma} + c_{{\bf k}\sigma}^{\dagger}d_{\alpha\sigma}
\right) +
\sum_{\bf{k},\sigma} \epsilon_{\bf k} c_{{\bf k}\sigma}^{\dagger} c_{{\bf k}\sigma},\nonumber
\end{eqnarray}
where the direct $U$ and $U'$ as well as exchange $J$ interactions between electrons of
spin $\sigma$ and on orbitals $\alpha$ or $\beta$ are taken into account. 

In the unrestricted Hartree--Fock approximation to the SIAM the local Green
functions $G^{\alpha\beta }_{\sigma}(\mu_{\alpha}, \mu_{\beta};\omega)^{HF}$
depend explicitly on the local 
magnetic moments $\mu_{\alpha}$ on
each orbital.
These local Green functions are used in the random phase approximation (RPA) to
determine the polarization diagrams
$\Pi^{\alpha\beta}_{\sigma\bar{\sigma}}(\mu_{\alpha}, \mu_{\beta}; \omega)$, corresponding to spin flip processes, and hence the
self--energies $\Sigma^{\alpha\beta}_{\sigma}(\mu_{\alpha}, \mu_{\beta};\omega)$.
The Green functions calculated in this way depend now on the local moments $\mu_{\alpha}$.
However, in a single impurity problem the spin-rotational symmetry cannot be
spontaneously broken.
To restore this symmetry within the approximate solution we use symmetrized
Green functions 
$
G^{\alpha\beta}_{\sigma}(\omega)=\frac{1}{2N_{\rm{orb}}} \sum_{\pm
  |\mu_{\alpha}|,\pm |\mu_{\beta}|}
G^{\alpha\beta}_{\sigma}(\mu_{\alpha}, \mu_{\beta};\omega),
$
where  $\pm |\mu_{\alpha}|$ in the sum means summation over different directions of
local orbital moments and $N_{orb}$ is the number of orbitals.
In this method the existence of the local moments is assumed from the beginning.

The lengths of the local moments $|\mu_{\alpha}|$ and
the number of particles on each orbital $n_{\alpha} $ are determined by
minimizing the ground state energy function, i.e.
$
E_{\rm{physical}} = \min_{\left\{\mu_{\alpha},n_{\alpha}\right\}} E(\mu_{\alpha},n_{\alpha}).
$
In this work we also adjust the Fermi energy of the
electrons in the impurity atom to have the proper total number of electrons. 

This generalized LMA is used to solve multi-orbital Hubbard
model $H_{\rm Hubb}=\sum_{ij}\sum_{\alpha,\sigma}t_{ij}^{\alpha}
d^{\dagger}_{i\alpha\sigma} d_{j\alpha\sigma}+H_{\rm local}$, where the local
part is a sum over all lattice site terms which are of the same form as the atomic part in  the SIAM. 
This model is solved within the DMFT where the self--consistency condition
relates the local matrix Green functions with the matrix of the
self--energies \cite{georges}. This condition simplifies for the Bethe lattice that is used
in this contribution. The band width $W_{\alpha}=2$ for each orbital $\alpha$ sets the energy units. 

\begin{figure}
     \centering
     \includegraphics[width=0.4\textwidth]{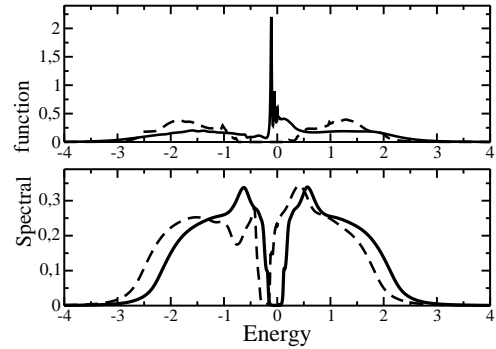}
     \caption{Spectral functions for two-orbital Hubbard model on Bethe lattice with infinite coordination number. Upper panel: degenerate case with $U=2.0$; solid line is for $J=0$, dashed line is for $J=U/4$. Lower panel: nondegenerate case with $|\epsilon_A-\epsilon_B|=0.3$,  $U=1.8$ and $J=U/4$; solid and dashed lines correspond to different orbitals. In both cases the Fermi level is at zero energy}
\label{fig1}
 \end{figure}

In Fig.~\ref{fig1} the spectral functions for two--orbital Hubbard model are
shown for degenerate case $\epsilon_A=\epsilon_B$ (upper panel) and for
nondegenerate (asymmetric) case $|\epsilon_A-\epsilon_B|=0.3$ (lower panel). 
Also the cases with ($J=U/4$) and without ($J=0$) Hund exchange coupling are
compared. The inter-band interaction $U'=U-2J$ is fixed preserving $SU(4)$
symmetry. In all cases the average number of electrons is fixed to $n_e=2$ per lattice site.


In the first case $\epsilon_A=\epsilon_B$ when $U$ is sufficiently large the
correlation gap is opened and the system is a Mott insulator. At smaller $U$,
when the system is on the metallic side, the Kondo--like peak is seen at $J=0$
and is destroyed at $J>0$.
When the orbital degeneracy is removed, as in $|\epsilon_A-\epsilon_B|=0.3$
case, the system for a given $U$ is metallic on one of the orbitals, whereas on
the other it is insulating. The number of electrons per site on each orbital
level is no longer equal and the spectral functions on different orbitals
begin to differ qualitatively. At larger $U$ the system is insulating in both
orbitals. 
So we find an orbital selective Mott-Hubbard MIT in asymmetric case. 
Increasing further the difference $|\epsilon_A-\epsilon_B|$ would lead to the 
appearance of the band insulator possessing a gap between bands.  
Initially the orbital selective Mott-Hubbard MIT was found in a two-orbital
symmetric case, where orbitals had the same energies but different
 bandwidths \cite{koga,liebsch,rice}. 
The present contribution extends this investigation on a case 
where this orbital degeneracy is removed. 
The crystal level splitting can be tuned in
some extend by the axial pressure and thereby there is a practical 
opportunity to search experimentally for the orbital selective MIT.

The generalized LMA to multi--orbital SIAM allows us
to efficiently study the correlated electron systems within DMFT. In particular
it is relatively easy to address the problems where the degeneracy of orbital
levels is removed, and hence the Fermi energy has to be determined
self--consistently in each iteration loop. 
Also the local moment approach is much more efficient in studying problems
with more then two orbitals \cite{kauch}. This regime is hardly accessible in numerically
exact approaches.

This work is supported by Polish Committee of Scientific Research through grant KBN-2 P03B 08 224. Partial support of the
Sonderforschungsbereich 484 of the Deutsche Forschungsgemeinschaft (DFG) is
also acknowledged.\\



\end{document}